# A FRAMEWORK FOR ENHANCING PERFORMANCE AND HANDLING RUN-TIME UNCERTAINTY IN SELF-ADAPTIVE SYSTEMS

Mohammed Abufouda

Department of Computer Science, Technical University of Kaiserslautern, Kaiserslautern, Germany

## ABSTRACT

Self-adaptivity allows software systems to autonomously adjust their behavior during run-time to reduce the cost complexities caused by manual maintenance. In this paper, a framework for building an external adaptation engine for self-adaptive software systems is proposed. In order to improve the quality of self-adaptive software systems, this research addresses two challenges in self-adaptive software systems. The first challenge is to provide better performance of the adaptation engine by managing the complexity of the adaptation space efficiently and the second challenge is handling run-time uncertainty that hinders the adaptation process. This research utilizes Case-based Reasoning as an adaptation engine along with utility functions for realizing the managed system's requirements.

## 1. INTRODUCTION

The majority of the existing work in the literature agrees [1] [2] that *self-adaptivity* in software systems is the ability of a software system to adjust its behaviour during run time to handle software system's complexity and maintenance costs [3] while preserving the requirement of the system. This property dictates the presence of an adaptation mechanism in order to build the logic of self-adaptivity without human intervention. Developing a self-adaptive software system is subjected to many challenges like handling the complexity of the adaptation space of the managed system. This complexity is conceived when the number of the states that the managed system can run in is relatively large. Also, this complexity manifests itself when new states are needed to be inferred from previous one i.e. learning from past experience. Another challenge is the uncertainty that hinders the adaption process during run-time. This paper will address these challenges. More precisely, our framework is concerned with the following problems:

- *Run-time uncertainty handling*: Uncertainty is a challenge that exists not only in self-adaptive software systems but also in the entire software engineering field on different levels. Therefore managing uncertainty is an essential issue in constructing a self-adaptive software system as uncertainty hinders the adaptation process if it is not handled and diminished.
- *Adaptation space:* The adaptation process raises a performance challenge if the adaptation space is relatively large, particularly when new adaptations are required to be inferred. This challenge requires an efficient mechanism that guarantees learning new adaptations as well as providing the adaptation with satisfactory performance. This means that the adaptation engine's response should be provided as soon as an adaptation is issued since late adaptations provided by the adaptation engine could be futile.

The rest of this paper is structured as follows: Section 2 lists the related work and the existing gabs in the literature. Section 3 shows a motivating example the will be used as a running example throughout this paper.In Section 4 we will demonstrate an overview of our solution while in Section 5 detailed information will be provided. Section 6 and 7contains the progress and the future of our research, in particular the evaluation. This paper concludes in Section 8.

## 2. RELATED WORK

The body of literature in the area of self-adaptivity has provided a plethora of frameworks, approaches and techniques to enhance self-adaptivity that is widespread in many fields. This section contains the related work to our research which has been introduced earlier as a short paper in [37]. In the following sections, we will present the related work categorized according to the mechanisms used to support self-adaptivity.

### 2.1 Learning based adaptation

Salehie and Tahvildari[2] proposed a framework for realizing the deciding process performed by an external adaptation engine. They used knowledge base to capture the managed system's information namely domain information, goals and utility information. This is used in the decision-making algorithm, as they name it, which is responsible for providing the adaptation decision. In [4], Kim and Park provided a reinforcement learning-based approach for architecture-based self-managed software using both on-line and off-line learning. FUSION [5] was proposed by Elkhodary et al. [5] to solve the problem of foreseeing the changes in environment, which hinders the adaptation during run time for feature-based systems using a machine learning technique. In [6], Mohamed-Hedi et al. provided a self-healing approach to enhance the reliability of web services. A simple experiment was used to validate their approach without empirical evidence.

### 2.2 Architecture & model based adaptation

RAINBOW [7] is a famous contribution in the area of self-adaptation based on architectural infrastructures reuse. RAINBOW monitors the managed system using abstract architectural models to detect any constraints violation. GRAF [8] was proposed for engineering self-adaptive software systems. The communication between the managed system and GRAF framework is carried out via interfaces. This approach has a performance overhead because GRAF reproduces a new adaptable version of the managed system. Similar to GRAF [8] Vogel and Giese [9] assumed that adaptation can be performed in two ways, parameter adaptation and structural adaptation. They provided three steps to resolve structural adaptation and used a self-healing web application as an example. Morin et al. [10] presented an architectural based approach for realizing software adaptivity using model-driven and aspect oriented techniques. The aim of this approach was to reduce the complexities of the system by providing architectural adaptation based solution. They provided model-oriented architectures and aspect models for feature designing and selection. Khakpour et al. [11] provided PobSAM, a model-based approach that is used to monitor, control and adapt the system behaviour using LTL to check the correctness of adaptation. Asadollahi et al. [12] presented StarMX framework for realizing self-management for Java-based applications. In their work they provided so called autonomic manager, which is an adaptation engine that encapsulates the adaptation logic. Adaptation logic was implemented by arbitrary policy-rule language. StarMX uses JMX and policy engines to enable self-management. Policies were used to represent the adaptation behaviour. This framework is restricted to Java-based application as the definition of processes is carried out by implementing certain Java interfaces in the policy manager. They evaluated their framework against some quality attribute. However, their evaluation for quality attributes was not quantified quantitatively. The work in [13] provided a new formal language for representing self-adaptivity for architecture-based self-adaptation. This language was used as an extension of the RAINBOW framework [7]. This work explains the use of this new language using an adaptation selection example that incorporates some stakeholders' interests in the selection process of the provided service which represents the adaptive service. Bontchev et al. [14] provides a software engine for adaptable process controlling and adaptable web-based delivered content. Their work reuses the functionality of the existing component in order to realize self-adaptivity in architecture-based systems. This work contains only the proposed solution and the implementation without experiment and evaluation.

## 2.3 Middleware based adaptation

In [15], a prototype for seat adaptation was provided. This prototype uses a middleware to support an adaptive behaviour. This approach was restricted to the seat adaptation which is controlled by a software system. Adapta framework [16] was presented as a middleware that enabled self-adaptivity for components in distributed applications. The monitoring service in Adapta monitored both hardware and software changes.

## 2.4 Fuzzy control based adaptation

Yang et al. [17] proposed a fuzzy-based self-adaptive software framework. The framework has three layers: (1) Adaptation logic layer, (2) Adaptable system layer, which is the managed system and (3) Software Bus. The adaptation logic layer represents the adaptation engine that includes the fuzzy rule-base, fuzzification and de-fuzzification components. This framework has a set of design steps in order to implement the adaptation. POISED [18] introduced a probabilistic approach for handling uncertainty in self-adaptive software systems by providing positive and negative impacts of uncertainty. An evaluation experiment had been applied which showed that POISED provided an accepted adaptation decision under uncertainty. The limitations of this approach are that it handles only internal uncertainty and does not memorize and utilize previous adaptation decisions.

## 2.5 Programming framework based adaptation

Narebdra et al. [19] proposed programming model and run time architecture for implementing adaptive service oriented. It was done via a middleware that solves the problem of static binding of services. The adaptation space in this work is limited to three situations that require adaptation of services. MOSES approach was proposed in the work [20] to provide self-adaptivity for SOA systems. The authors used linear programming problem for formulating and solving the adaptivity problem as a model-based framework. MOSES aimed to improve the QoS for SOA, and the work in [20] provides a numerical experiment to test their approach. QoSMOS[21] provided a tool-supported framework to improve the QoS for the service based systems in adaptive and predictive manner. The work in [22] provided an implementation of architecture-based self-adaptive software using aspect oriented programming. They used a web-based system as an experiment to test their implementation. Their experiment showed that the response time of the self-adaptive implementation is better than the original implementation without a self-adaptivity mechanism. Liu and Parashar[23] provided Accord, which is a programming framework that facilitates realizing self-adaptivity in self-managed applications. The usage of this framework was illustrated using forest fire management application.

Table 1, which is similar to what proposed in [24], summarizes the related work done in this research. The table has two aspects of comparison (1) Research aspects and (2) Self-adaptivity aspect. The earlier aspect is important and represents an indication regarding the maturity and creditability of the research. The later aspect is related to the topic of this paper.

| Covered literature categorization | Work | Research aspects | | | | | | Self-adaptive software system aspects | | | | | |
|---|---|---|---|---|---|---|---|---|---|---|---|---|---|
| | | Problem Statement | Contribution statement | Experiment | evaluation metrics | Limitations | Threats to validity | Adaptation Expediency | Adaptation remembrance | Uncertainty Handling | Adaptation Res. Time | Adaptation style | Adaptation engine |
| Learning based adaptation | [2] | √ | √ | X | X | X | X | X | X | X | X | Dynamic | External |
| | [4] | √ | √ | √ | X | X | X | X | X | X | X | Dynamic | External |
| | [5] | √ | √ | √ | √ | √ | X | √ | X | X | √ | Dynamic | External |
| | [6] | X | X | √ | X | X | X | X | X | X | X | Dynamic | External |
| Architecture & model based adaptation | [7] | √ | √ | √ | X | X | X | X | X | √ | X | Dynamic | External |
| | [8] | √ | √ | √ | X | X | √ | X | X | X | X | Dynamic | External |
| | [9] | √ | √ | √ | X | X | X | X | X | X | X | Static | Internal |
| | [12] | X | X | √ | X | X | X | √ | X | X | X | Dynamic | External |
| | [10] | X | X | √ | X | X | X | √ | X | X | X | Dynamic | Internal |
| | [11] | √ | X | X | X | X | X | X | X | X | X | Dynamic | Internal |
| | [13] | √ | √ | X | X | X | X | X | X | X | X | Static | External |
| | [14] | √ | √ | X | X | X | X | √ | X | X | X | Dynamic | External |
| Middleware based adaptation | [15] | √ | √ | √ | X | X | X | X | X | X | X | Static | Internal |
| | [16] | √ | √ | X | X | X | X | X | X | X | X | Dynamic | External |
| Fuzzy control based adaptation | [17] | √ | √ | X | X | X | X | X | X | X | X | Dynamic | External |
| | [18] | √ | √ | √ | X | X | X | √ | X | √ | √ | Dynamic | Internal |
| Programming framework based adaptation | [19] | X | X | √ | √ | X | X | X | X | X | X | Dynamic | External |
| | [20] | √ | √ | √ | X | X | X | X | X | X | X | Dynamic | External |
| | [22] | √ | √ | √ | √ | X | X | √ | X | X | √ | Dynamic | Internal |
| | [23] | √ | √ | √ | X | X | X | √ | X | X | √ | Dynamic | Internal |

Table 1: Summary of related work

# 3. MOTIVATING EXAMPLE

The motivating example is a software system controlling a robot that requires self-adaptive behaviour during run-time. This motivating example is used for both motivating the need for self-adaptive software systems and for the experimentation and the validation of our framework. The idea of the robot is derived from [18] with an attribute extension for a more variety of configurations. Figure 1 shows an abstract view of the robot managed system which has an exploratory task and should submit the captured videos to a remote controlling centre. Even though the example is from the robotics field, we emphasize that our concern is only the software system that manages the self-adaptive behaviour of the robot rather than the robot itself. This means that the robot as a managed system could be any other system that requires enabling the self-adaptation property. We will use this example as a running example through this paper.

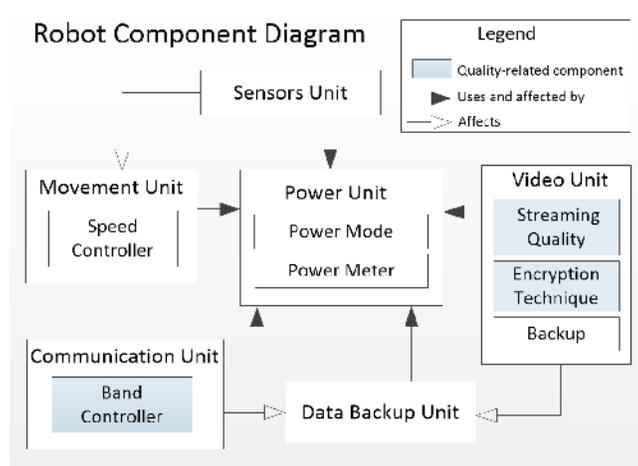

Figure 1: Robot Components

The components in Figure 1 are dependent on each other; one component may affect other component(s). This dependency contributes to providing a set of various possible states of the robot, which is useful in explaining how our framework works. The robot requires adaptation of its behaviour during run-time in order to keep fulfilling its requirements without manual controlling from the remote controlling centre. This adaptation is a response to the changes in the environment where the robot is working and/or the changes in the attributes of the robot itself e.g. the speed and the power. These requirements are quality of service (QoS) requirements and functionality requirements that need to be achieved by the robot self-adaptively.

An example of QoS requirements is *Video Quality* where the robot aims at keeping the quality of the transmitted video as good as possible. This is done by selecting the appropriate video quality automatically during run-time. The available power affects this requirement because higher video qualities require more power consumption than lower ones. The robot should control this process efficiently. Another example of QoS requirements is *Transmission Security* where the robot should keep the transmitted data as secure as possible during submitting it to the remote controlling centre. This is achieved by selecting one among a set of encryption techniques where each technique has its advantages and drawbacks in terms of power consumption, security level, and encryption performance. An example of functionality requirements is *Robot Fitness* where the robot should manage the relations among its attributes in order to keep itself as fit as possible. For instance, the robot should reduce its speed if the power is not sufficient or an obstacle is detected by the sensors unit. Another example of functionality requirements is to enable the data backup if the communication with the remote centre is lost. This requires choosing a suitable video quality due to the limitation of the space of backup storage.

The challenges that the robot system may face in the self-adaptation context and are addressed by our framework automatically are:

- *Run-time uncertainty handling*: The robot may fail to identify one of its environmental variable values during its operation. For example the sensors may fail to tell whether there is an obstacle in the area or not. In such problematic situations, the robot should behave tolerably; otherwise the robot may run into unwanted states.

- *Adaptation space complexity impacts*: If the robot has $N$ attributes each of them has $M$ different possible values, then the possible states S that the robot may run in are: $\prod_{i=1}^{N} Mi$ This requires an efficient handling of these operating states that guarantees accepted performance. Concretely, the response time of the adaptation engine is a crucial issue because the delayed adaptation response could be useless. For example, if the robot's communication with the remote centre has been lost, then the robot should start the back-up storage in order to keep all the captured videos. Such decisions should be provided to the robot immediately; otherwise the robot could deviate from its requirements.

## 4. SOLUTION OVERVIEW

In this section, an overview of the solution will be presented. Based on Figure 2, which illustrates framework reference model, the following subsections describe the Managed system and adaptation engine that is decomposed into the Adaptation mediator and the Case-based reasoning engine.

### 4.1 The managed system

The managed system is the system that needs to adapt its run-time behaviour autonomously e.g. the robot system discussed in Section 2. The managed system must provide a set of its self-adaptation concerned attributes. An example of these attributes, based on the motivating example discussed in Section 2, is shown in Table 2. The table also shows the complexity of the adaptation space size i.e. the robot may run in one of 8640 possible different configurations.

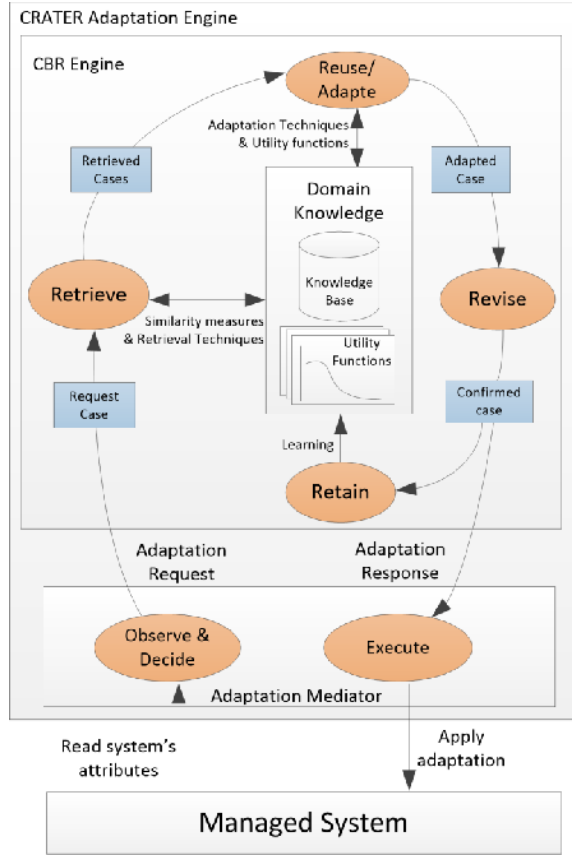

Figure 2: Framework Reference Model

| Attribute | Values set |
|---|---|
| Communication | OFF, VHF, X-band, UHF |
| Power Mode | Full Power, Medium Power, Saving Mode |
| Power Meter | Low, Medium, High |
| Speed | Low, Medium, High |
| Video quality | Very low, Low, Medium, High, Very High |
| Data Backup | On, Off |
| Obstacles | True, False |
| Encryption | Zig-Zag Permutation, Puer Permutation, Naive, Video Encryption Algorithm |

Table 2: Robot Attribute Data Sheet

## 4.2 Adaptation engine

This section provides details about the components of the adaptation engine.

### 4.2.1 Adaptation Sample

Before digging deeper in the model's details, it is better to show how our solution works with two adaptation samples. We assume that the managed system, the robot in our case, provides a service utility U and an adaptation process is issued when this utility is below or is approaching 1 a predefined utility threshold UT. Table 3 illustrates two randomly selected adaptations from the experiment that will be discussed later, one of them contains uncertain value. The first adaptation request embraces a defect in the operating mode of the robot as there is an obstacle while the robot speed is high which represents a violation. The adaptation response for this unwanted state of the robot is to reduce the speed. Reducing the speed is the only possible adaptation response as we cannot change the obstacle to false as it is not adaptable attribute. The table shows that the utility of the adaptation request is 0.484 which is a utility threshold breaker, assuming that UT is 0.5.

The adaptation engine managed to provide an adaptation response with utility 0.892 which is greater than 0.5. The other adaptation requests hold uncertain value in the communication attribute. The adaptation engine issued adaptation process for this robot state because the uncertain attribute, the communication, is uncertain and one possible values, off, leads to utility less than UT. When the communication attribute goes off, it breaks the UT, which means that the robot is unable to establish a connection with the remote centre. As a result an adaptation process is issued that produces the adaptation response that assures that the communication is set with appropriate value to enable communication with the remote centre. Needless to say that the chosen value, UHF, should not break the utility of the robot which is satisfied and the utility is 0.8666. Another possible adaptation response for the second adaptation request is to enable the data back up and to set off the communication. However, the adaptation engine did not choose this scenario because its utility is less than the utility of the chosen adaptation response. This is because the ultimate goal of the framework is to maximize the utility of the managed system.

| Attribute | Ad.Req.1 | Ad.Res.1 | Ad.Req.2 | Ad.Res.2 |
|---|---|---|---|---|
| Communication | UHF | UHF | ? | UHF |
| Power Mode | Saving Mode | Saving Mode | Medium Power | Medium Power |
| Power Meter | High | High | High | High |
| Speed | High | Low | Low | Low |
| Video quality | Very High | High | Low | Low |
| Data Backup | Off | Off | Off | Off |
| Obstacles | True | True | False | False |
| Encryption | Pure Perm. | Puer Perm. | Zig-Zag Perm. | Zig-ZagPermu. |
| Utility | 0.484 | 0.892 | ? | 0.8666 |

Table 3: Adaptation Sample

### 4.2.2 Adaptation mediator

Now, we can present in more details the description of our framework. As shown in Figure 2, the adaptation mediator is responsible for:

- *Monitoring* the managed system by reading its attributes to decide whether an adaptation is required or not. The framework expects that the managed system provides a service with overall utility *U*. The *adaptation request* is the set of attributes' values of the managed system at the time of issuing the adaptation process. Consequently, the adaptation request is sent to the adaptation engine to start the adaptation process.

---

[1]This is because our solution treats self-adaptation in a reactive or a proactive way depending on the implementation of monitoring process within the adaptation mediator.

- *Executing* the adaptation response received from the adaptation engine. The adaptation response is the result of the adaptation process performed by the adaptation engine, which is the corrective state to be applied on the managed system.

### 4.2.3 Case-based reasoning engine

The adaptation engine is built mainly on Case-based Reasoning (CBR) which facilitates the automation process of the adaptation. CBR is an artificial intelligence paradigm that mimics the human behaviour in solving problems based on the solutions of previous and similar problems. Generally, a case is an object that encapsulates some attributes e.g. the robot attributes shown in Table 3 and, traditionally, the attributes of a case are divided into problem related attributes and solution related attributes. In our work we model the adaptation request as problem part of a CBR case and the adaptation response as solution part of a CBR case. Specifically, the red attributes in Table 3 represents a problem part of a CBR case and the green attributes represents the solution part of a case. The task of our framework is to find out an appropriate solution for these red attributes. Traditional CBR life cycle, as shown in Figure 3, consists of four stages:

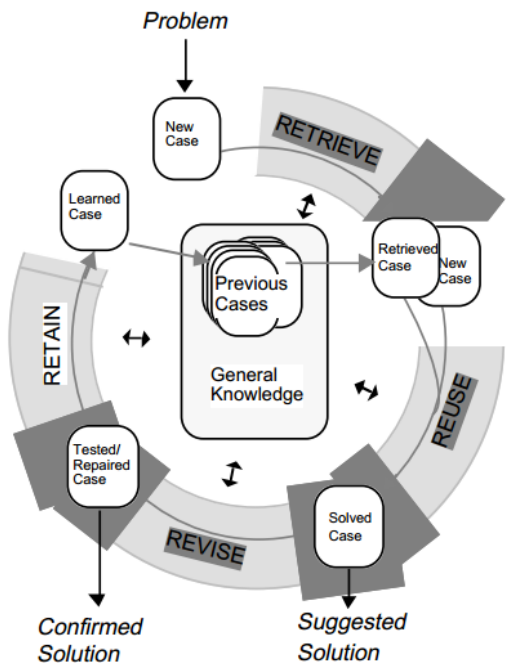

Figure 3: Case-based Reasoning Life Cycle [25]

- *Retrieve*: The CBR system retrieves the most similar case(s) from the Knowledge Base by applying the similarity measures on the request case. In [25] [26] [27], many similarity measures for improved case retrieval have been introduced. Figure 4 shows an example of how the similarity is performed on the cases from the knowledge base and which attributes are considered in the similarity measures.

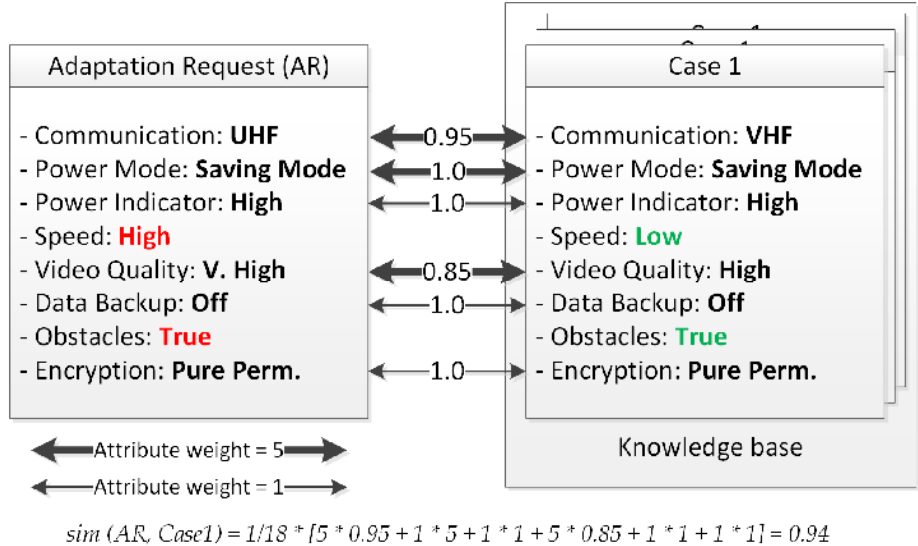

Figure 4: Example of similarity measure between adaptation request and a case from the

- *Reuse (Adapt):* In this stage, CBR benefits from the information of the retrieved cases. If the retrieved cases are not sufficient in themselves to solve the request case, the CBR engine adapts this/these case/s to generate a new solution. Some of the common techniques for reusing and adapting the retrieved knowledge are introduced in [28]. We use Generative Adaptation [29], which requires some heuristics, e.g. utility functions, to provide an efficient adaptation process.
- *Revise*: A revision of the new solution is important to make sure that it satisfies the requirements of the managed system. The revision process can be done by applying the adaptation response to real world, evaluate it by the domain expert, or by simulation approaches. To enhance the automation of the adaptation process, we use utility functions which revise the generated adaptation and judge its utility satisfaction on the fly.
- *Retain*: In this stage, the new generated cases are saved in the knowledge base. Case-Based Learning (CBL) has been introduced in [30] to provide algorithms and approaches for an efficient retain process.

# 5.    MODEL AND SPECIFICATIONS

In this section we explain how we tackle the challenges described in Section 1. Precisely, it explains the adaptation process and how the utility functions are used.

## 5.1 The knowledge base

The knowledge base in our framework contains the states of the managed system that satisfy its requirements. This property is guaranteed in the retain process where no case is retained unless it has a utility greater than UT. The knowledge base is modelled by the domain experts by capturing all attributes of the managed system that are related to the adaptation process. The operations performed on the knowledge base are restricted to case retrieval and case retention. However, the domain expert could investigate it for offline maintenance e.g. adds new cases, remove cases and alter cases. Table 4 shows an excerpt from the knowledge base for the motivating example discussed in Section 2. Assuming that the utility threshold is 0.5, it is clear from the table that all the cases in the knowledge base have a utility greater than the utility threshold.

| Attribute | C1 | C2 | C3 | C4 | C5 |
|---|---|---|---|---|---|
| Communication | UHF | VHF | VHF | UHF | UHF |
| Power Mode | Medium | Medium | Full | Full | Medium |
| Power Meter | High | High | High | Low | High |
| Speed | Low | Medium | Medium | Medium | Medium |
| Video quality | V.Low | High | V.High | Medium | Medium |
| Data Backup | Off | Off | Off | On | Off |
| Obstacles | False | False | False | True | True |
| Encryption | PuerPermu. | Zig-Zag Perm. | VEA | Puer Perm. | VEA |
| Utility | 0.813 | 0.603 | 0.758 | 0.565 | 0.928 |

Table 4: Excerpt from the knowledge bass

## 5.2 The managed system attributes

The managed system operating states are modelled as CBR cases. Each case has a set of attributes and each attribute has a type and a weight.

### 5.2.1 Attribute types

Case attributes can be flagged as one or more of the types shown in Table 5. During the design of the managed system, each attribute must be labelled as adaptable or unadaptable. During the analysis process of the adaptation request, we identify UT-breaker and utility-antagonist attributes. The framework alters the UT-breaker to provide adaptation response with utility greater than the UT. For providing an optimal adaptation response (Optimization problem), the utility-antagonist attributes is altered, which raises the utility of the provided adaptation response.

| Attribute Type | Description |
|---|---|
| Adaptable | An attribute whose value can be changed during the adaptation process e.g. Speed. |
| Unadaptable | An attribute whose value cannot be changed during the adaptation process e.g. Obstacles |
| UT-breaker | An attributes whose value participates in reaching a goal-violating state. |
| Utility-antagonist | An attribute whose value participates in decreasing the overall utility. |

Table 5: Managed System Attributes Types

### 5.2.2 Attributes weights

It is normal that the attributes of the managed system vary in their effect on the utility of the provided service. Based on that, Pareto principle is applied and each attribute is weighted in order to provide optimal representation of the state of the managed system.

## 5.3 Utility functions

Utility functions are incorporated in the reference model in order to: (1) assess the cases of the knowledge base in terms of satisfying the requirements of the managed system, (2) provide a heuristic for the adaptation process and provide affirmation regarding the adaptation response expediency, (3) analyse the adaptation requests to identify UT-breaker attributes and (4) determine when to issue the adaptation process; i.e. if the managed system's overall utility reaches or is approaching the UT.

### 5.3.1 Utility function definition

Utility function is a function that maps a set of attributes to a value if certain condition holds. For simplicity, the utility function definition is based on the work in [31] and extended in order to combine multiple utility-involved attributes.

The utility function is defined as in Equation 1:

$$Utility_{(a_1,...,a_i)} = \begin{cases} v_1 & if\ condition_1\ holds \\ v_2 & if\ condition_2\ holds \\ . \\ . \\ v_{n-1} & if\ condition_{n-1}\ holds \\ v_n & Otherwise \end{cases}$$

*Equation 1*

where:

- $(a1,...,ai)$ is the set of involved managed system attributes.
- $(v1,...,vn)$ are the values of the utility function.
- $(condition_1 ,..., condition_{i-1})$ is a set of condition for satisfying the utility function.

An example of the utility function is shown in Equation 2 which describes the relation among *Power Mode*, *Video Quality* and *Encryption Technique*:

$$U_{(P,Q,E)} = \begin{cases} 0.1 & if\ (P=3\ and\ (Q=1\ or\ Q=2)\ and\ E=1)\ holds \\ 0.5 & if\ (P=2\ and\ Q=2\ and\ E=1)\ holds \\ 0.8 & if\ (P=1\ and\ Q=3\ and\ E=3)\ holds \\ 0.99 & Otherwise \end{cases}$$

Equation 2

### 5.3.2 Utility function weight

In reality, the adaptation-involved attributes of the managed system can be shared by more than one utility function due to the correlation among these attributes. Weighting these utility functions is a crucial issue in modelling the managed system's requirements. The weighting process is normally the task of the domain expert and can be improved by weight learning.

### 5.3.3 Overall utility function

The {Weighted Geometric Mean} (WGM) is used to estimate the overall utility of the managed system in terms of its utility functions. If we have a set of utility function values U= {$u_1,u_2,...,u_n$} with corresponding weights W={$w_1,w_2,...,w_n$ }, then the overall utility is estimated by the following equation:

$$U_{overall} = (\prod_{i=1}^{n} u_i^{w_i})^{1/(\sum_{i=1}^{n} w_i)}$$

Equation 3

## 5.4 Adaptation process

In this section we describe the adaptation process shown in Figure 5. The adaptation process goes through the following phases:

### 5.4.1 Analysing adaptation request

When the adaptation engine receives an adaptation request, it analyzes it to identify the attributes that breaks UT and the attributes that antagonize the managed system utility. This identification process is done by comparing the adaptation request values to the utility functions. That is any attribute participates in making any of these utility function below the UT is considered as utility breaker attribute. Similarly, any attribute decreases any of the utility functions is considered as antagonistic attribute. The identification of these attributes helps in providing efficient adaptation response by changing the values of these two types of attributes to get higher utility from the adaptation response.

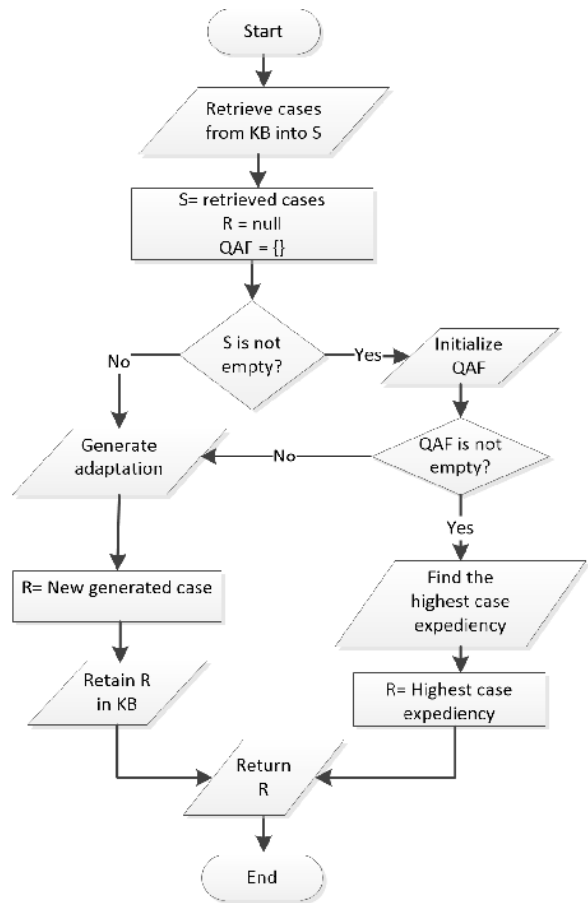

Figure 5: Adaptation Process Flow Chart

**5.4.2 Case retrieval**

Case retrieval is CBR core functionality. We retrieve the most similar case(s), if any, to the *request case* as shown in Figure 2. It is important to mention that, the request case is formulated from the adaptation request by excluding the UT-breaker attributes from it. This exclusion is inevitable as the knowledge base keeps only cases of best operating states that have no UT-breaker values at all. After this formulation of the request case, it is ready for the similarity measure calculation, as shown in Figure 4, to find its best matching cases. For example, if the robot system is running with the following attributes: *{Power= Full power, Video Quality= Very High, Obstacles=True, Speed= Fast}*, then, obviously, this state represents an unwanted operating state because the robot speed should not be fast if an obstacle is detected. This state is a typical adaptation request as it represents a deviation from the system requirements that are defined by a utility function similar to Equation 2. As the speed is the UT-breaker attribute in this example, the request case will be formulated by excluding it from the adaptation request. So the adaptation case of that adaption request is *{Power= Full power, Video Quality= Very High, Obstacles=True}*.

**5.4.2 Constructing qualified adaptation frame (QAF)**

The retrieval process for adaptation requests returns a set of cases called *Qualified Adaptation Frame*(QAF), such that each case $C_k$in this set satisfies the condition:

$$\beta \leq Sim(C_k, Adap_{Req}) \leq 1$$

Equation 4

Where  is a value between [0,1] and represents the minimal similarity value for accepting retrieved cases from the knowledge base and *Sim* is a function that calculates the similarity between the adaptation case and each of the retrieved case.Having the *QAF* ready, a decision on which adaptation response to select has to be taken. In fact, similarity is not the only decisive factor, however, the utility of the retrieved case is also considered. This combination is called *Case Usefulness*. We calculate the usefulness of a case in the QAF by Equation 5:

$$CU(c)_{QAF} = 1 - [(1 - sim(Adap_{req}, c)) * utility(c)]$$

Equation 5

Where*CU* is the Case *Usefulness* for a case *c* in the QAF and *sim* is the similarity between the adaptation request $Adap_{req}$ and the case *c*. This novel combination in calculating case usefulness is essential. On the one hand, the inclusion of similarity of the retrieved cases in calculating case usefulness is important as higher similarity leads to fewer changes in the managed system attributes. On the other hand, the inclusion of the utility reflects the quality of the case in terms of meeting the managed system's requirements.

**5.4.3      Generating adaptation response**

If the QAF is empty, the adaptation response is generated based on the utility function by adapting the request case attributes in order to provide a case with a utility greater than UT. This process is called *Utility-guided constructive adaptation*which has two flavours. (1)First Fit Heuristic: This is a normal iterative search process in the space values of the attributes that is applied on the request case [32]. The first value that causes the utility of adaptation request to be greater than UT is returned as an adaptation response.(2) Best Fit Heuristic: which is an extension of the first fit heuristic with extra capability; that is the search process finds values that maximize the utility of the adaptation response (i.e. providing an optimal adaptation). If the adaptation response is generated by one of the previous ways, the utility of the generated case is considered as the case usefulness.

### 5.4.4 Retaining

Retain phase is restricted to the newly generated adaptation response from the Utility-guided constructive adaptation process. As all of the generated adaptation responses have a utility greater than UT, they are qualified for retaining in the knowledge base for future reuse.It is clear that our model is able to start operating with an empty knowledge base, which enables a full automation of the adaptation process. The utility functions govern the learning process, which guarantees the quality of retained cases. The number of retained cases in the knowledge base decreases overtime which raises the likelihood of retrieving the adaptation response instead of generating it. This has a positive impact on the performance and reduces the response time of the adaptation engine significantly. Algorithm 1 abstracts the automation of adaptation process of our solution.

---

**Algorithm 1** Adaptation process

**Require:** $KB$ , $A_{req}$
**Ensure:** $Utility(A_{res}) > UT$
1: $List\ cases \Leftarrow Retrieve\ (KB, A_{req})$
2: $List\ QAF$
3: $Case\ A_{res}$
4: **while** $Case\ c \Leftarrow Iterate(cases)$ **do**
5:    **if** $Sim(A_{req}, c) \in [\beta, 1]$ **then**
6:       $QAF.add(c)$
7:    **end if**
8: **end while**
9: **if** $QAF$ is not $Empty$ **then**
10:    $A_{res} \Leftarrow max(CaseExpediency(QAF)\ )$
11:    **Return** $A_{res}$
12: **else**
13:    $A_{res} \Leftarrow ConstructiveAdapt(A_{req})$
14:    **Retain**$(A_{res}, KB)$
15:    **Return** $A_{res}$
16: **end if**

---

Algorithm 1

### 5.5 Run-time uncertainty diminution

Our framework's ultimategoal is to provide an adaptation response that maximizes the utility of the managed system. Therefore, when the managed system is running under uncertain state, consequently its utility is not deterministic;we need to quantify this uncertainty to provide efficient adaptation responses. To that end, we are identifying uncertainty by capturing its three dimensions [28]:

- *Location* of uncertainty: uncertainty is revealed within our model in two locations. The first location (Location 1) is the managed system state and second location (Location 2) is within the QAF.

- *Nature* of the uncertainty in Location 1 is the run-time uncertainty which is the knowledge shortage in the managed system attributes' values. This could be due to environmental reasons or measurement errors[2] in providing known values. The nature of uncertainty in Location 2 is the variability. This means that the QAF has more than one case with the same maximum highest usefulness.

- *Level* of uncertainty needs to be estimated. Otherwise, we will not be able to decide whether an adaptation is required or not. To estimate the level of uncertainty in Location 1, we start with generating a set □ of all possible states that the uncertain state can be one of them. Then, the number $\Re$ of states that belongs to □ and require adaptation is calculated. Subsequently, the probability $\mu$ that the uncertain state is a UT-breaker is determined as seen in Equation 6. Also, the uncertainty degree in the managed system is estimated as shown in Equation 7.

$$\mu = \frac{\Re}{Size(\kappa)}, \ \Theta = \frac{\#uncertain\ attributes}{\#all\ state\ attributes}$$

Equation 6

Finally, the overall uncertainty $\eta$ *Level* is estimated by Equation 7.

$$\eta = 1 - [(1 - \mu) * (1 - \Theta)]$$

Equation 7

Even though our work handles the run-time uncertainty in Location 1, by calculating the uncertainty level, our framework provides a naive solution for estimating the level of the uncertainty in Location 2 (due to variability in the QAF). This solution does not require any further calculations. That is, if there is more than one case with the same highest usefulness in the QAF, then the selected case is the case with the highest utility as it satisfies the ultimate goal of the framework we mentioned earlier in this section.

In the context of utility functions, there are two ways to deal with uncertainty: (1) *Optimistic Paradigm*: which deals with the uncertain values as values that heighten the utility and (2) *Pessimistic Paradigm:* which deals with the uncertain values as values that belittle the utility. Both pessimistic and optimistic paradigms are not preferable in systems like the one in our motivating example. This is because when the robot is operating in an optimistic paradigm and its uncertain state dictates an adaptation, the optimistic paradigm will fail to issue an adaptation process. Likewise, when the robot is operating in a pessimistic paradigm and its uncertain state do not dictate an adaptation; the pessimistic paradigm will cause performance overhead for issuing useless adaptation. To diminish the run-time uncertainty efficiently, we introduce a *Hybrid Paradigm* that depends on a cut-off value, $_{threshold}$[3]. If $_{threshold}$ is *one* then it behaves pessimistically i.e. an adaptation process is issued whenever the managed system runs in uncertain state, and when $_{threshold}$ is *zero*, it behaves optimistically i.e. no adaptation process is issued. Intuitively, $_{threshold}$ should maintain a value greater than zero and less than one. An adaptation process is issued only when is less than or equal $_{threshold}$.

---

[2] For example sensor or actuator errors and problems
[3] This value is defined during configurations time

# 6. PROGRESS AND CURRENT STATUS

A prototypical implementation of the solution has been done. This implementation includes the integration of the CBR engine with utility functions. The implementation also includes the generative adaptation of the adaptation responses. Moreover, uncertainty analysis and quantification are provided in this implementation paving the way for handling uncertainty during run-time. The three dimensions of the uncertainty [28] has been modelled and implemented.

# 7. FUTURE DIRECTION AND EVALUATION

For future direction, firstly, we will use a case study to empirically evaluate and validate our approach. The case study i.e. the managed system should require the self-adaptivity mechanism that performs well under run-time uncertainty. Secondly, we will evaluate the results of the case study application. The evaluation will be based on software quality metrics and GQM [29]. We expect that the experimentation of our solution will provide a positive potential results for both handling the uncertainty and the complexity of adaptation space. However, we do not have a clue regarding the response time of the adaptation engine, the results will reveal this issue.

# 8. CONCLUSION

In this paper, we have presented our theoretical approach for realizing self-adaptivity in software systems. We started by showing the gabs in the research and the expected contributions of the research. Also, we have presented details about the solution model and the used technology, Case-based reasoning. The progress of the work was presented along with the future directions. This paper ended with our vision of the evaluation process of our solution.

## Authors

**Mohammed Abufouda** received the BSc. in Computer Engineering from Islamic University, Palestine in 2006 and the MSc. degree in computer science from Technical University of Kaiserslautern, Germany, in 2013. He is a PhD candidate at Technical University of Kaiserslautern in computer science department. He is IEEE and ACM student member and his research interests include software engineering and complex system analysis.